\documentclass[twocolumn,titlepage,preprintnumbers,amsmath,amssymb,aps,prl,superscriptaddress,reprint]{revtex4-1}

\usepackage{graphicx}
\usepackage{color}
\usepackage{upgreek}
\usepackage{pgffor}
\usepackage[caption=false]{subfig}

\newcommand{\sampt}[2]{#1\ensuremath{_\textrm{#2}}}
\newcommand{\mw}{\ensuremath{M_\textrm{w}}}
\newcommand{\mn}{\ensuremath{M_\textrm{n}}}
\newcommand{\mc}{\ensuremath{M_\textrm{c}}}

\setcitestyle{super}

\begin{document}

\title{Short Chains Enhance Slip of Highly Entangled Polystyrenes during Thin Film Dewetting}

 \author{S. Mostafa Sabzevari }
\affiliation{Department of Mechanical and Industrial Engineering, Concordia University, Montreal, Canada}
\author{Joshua D. McGraw}
\affiliation{Soft Matter Physics Group, Experimental Physics, Saarland University, 66041 Saarbr\"ucken, Germany}
\affiliation{D\'epartement de Physique, Ecole Normale Sup\'erieure / PSL Research 
University, CNRS, 24 rue Lhomond, 75005 Paris, France}
  \author{Paula Wood-Adams}
\affiliation{Department of Mechanical and Industrial Engineering, Concordia University, Montreal, Canada}

 \begin{abstract}
We investigate the effect of short chains on slip of highly entangled polystyrenes (PS) during thin film dewetting from non-wetting fluorinated surfaces. Binary and ternary mixtures were prepared from monodisperse PS with weight average molecular weights $5 < \mw < 490$\,kg/mol. Flow dynamics and rim morphology of dewetting holes were captured using optical and atomic force microscopy. Slip properties are assessed in the framework of hydrodynamic models describing the rim height profile of dewetting holes. We show that short chains with \mw\ below the polymer critical molecular weight for entanglements, \mc, can play an important role in slip of highly entangled polymers. Among mixtures of the same \mw, those containing chains with $M<\mc$ exhibit larger slip lengths as the number average molecular weight, \mn, decreases. The slip enhancement effect is only applicable when chains with $M<\mc$ are mixed with highly entangled chains such that the content of the long chain component, $\phi_\textrm{L}$, is dominant ($\phi_\textrm{L}<0.5$). These results suggest that short chains affect slip of highly entangled polymers on non-wetting surfaces due to the physical or chemical disparities of end groups, and any associated dynamical effect their presence may have, as compared to the backbone units. The enhanced slip in this regard is attributed to the impact of chain end groups or short chain enrichment on the effective interfacial friction coefficient. Accordingly, for entangled PS, a higher concentration of end groups or short chains at the interface results in a lower effective friction coefficient which consequently enhances the slip length.
\end{abstract}

\maketitle

In contrast to the commonly applied no-slip boundary condition of small molecule fluids, wherein the bulk fluid flow extrapolates to zero velocity at the solid boundary, polymer melts can exhibit a considerable range of boundary conditions from no slip to strong slip. The extent of polymer slip depends on a variety of factors, mainly including the substrate surface chemistry and physical structure, and the polymer chain size.\cite{archer05TXT} Highly entangled polymer melts exhibit strong slip during flow on low surface energy substrates, however, the slip length is significantly reduced if highly adsorbing substrates are used instead.\cite{archer05TXT} Aside from the fundamental interest associated with polymer molecules flowing against a solid boundary, a thorough understanding of the extent of polymer slip is also critical in many practical applications. Knowledge of slip properties allows for the correct measurement and characterization of material properties such as viscosity,\cite{morrison01TXT} as well as allowing for improved manufacturing processes,\cite{legrand98jr, Ghanta} spreading of paints and coatings,\cite{PGG85rmp} and surface finishes of polymeric products.\cite{Denn} Such knowledge is also vital for developing more accurate microscopic slip theories for better model predictions.\cite{BrochardPGG, Tchesnokov} 

In polymer flow, slip can be due to cohesive and/or adhesive failure, the failure mode is mainly governed by the substrate surface energy and shear rate.\cite{Denn, BrochardPGG, Tchesnokov, Brochard94LMR} On a high surface energy substrate such as glass or steel, some polymer chains adsorb onto the solid wall forming an adsorbed layer of molecular size above the solid/liquid interface while others move with the bulk flow above them. During flow at shear rates large compared to molecular relaxation frequencies,\cite{BrochardPGG, Leger97} the polymer starts to slip away on the adsorbed layer due to cohesive failure. Slip in this case is called apparent slip since the chains in direct contact with the substrate are not in motion, yet the bulk velocity profile does not extrapolate to zero at the substrate. At high shear rates, the slip mechanism may change to adhesive failure if desorption occurs.\cite{Brochard94LMR} Dynamics of slip on high surface energy substrates is therefore mainly controlled by the density of adsorbed chains, their bonding energy to the wall, and the number of entanglements between adsorbed and mobile chains.\cite{Brochard94LMR, Leger97, Sabzevari14M, Sabzevari14M2} 

On smooth and non-adsorbing surfaces,\cite{Brochard94LMR} for example substrates with low surface energies, slip is more commonly attributed to adhesive failure with mobile chains directly in contact with the solid surface. Polymer slip on non-adsorbing walls, called true slip, is shear stress independent in the Navier linear model and controlled by the chain/wall friction coefficient and polymer viscosity.\cite{BrochardPGG, Fetzer07L} A mesoscopic inference of the flow profile, as discussed here for the case of dewetting polymer films, can lead to a determined slip length which is a combination of true and apparent slip. 

The slip behavior of monodisperse polymers on high surface energy\cite{archer05TXT, Leger97} and low surface energy substrates\cite{Fetzer07L, Baumchen12} has been studied extensively in the past since monodisperse melts are considerably simpler to model. Industrial polymers, however, are polydisperse largely due to cost considerations with regard to polymerization of monodisperse polymers.\cite{Hatzikiriakos12} The polydispersity effect on apparent slip of polymers has previously been explored.\cite{Sabzevari14M, Sabzevari14M2, Hatzikiriakos12, Ansari1, Ansari2, inn} In these studies, it was shown that the polymer slip on adsorbing walls may significantly be affected by the size and content of short chains in the system. Interestingly, the presence of weakly entangled chains in an otherwise highly entangled melt can in fact enhance the slip length.\cite{Sabzevari14M, Sabzevari14M2}. This enhanced slip was partly attributed to the enrichment of short chains at the slip plane which reduces the number of entanglements between the adsorbed and mobile chains.\cite{inn, Sabzevari3} 

The effect of polydispersity on slip near low surface energy substrates has not yet been fully explored. Theoretical\cite{BrochardPGG, Brochard94LMR, Brochard96MAC} and experimental\cite{Baumchen12} studies on slip of polymers on solid substrates have, to the best of our knowledge, been focused on monodisperse samples only. Polymer melts with large polydispersity indices are rich in the number of end groups but may also contain a significant number of entanglements. While entanglements are known to enhance slip through an enhanced viscosity as compared to the monomeric friction coefficient, chain ends as well as short chains are known to localize near interfaces\cite{Tanaka1, Tanaka2, Minnikanti} as compared to the bulk, with subsequent measurable effects on surface and interfacial properties,\cite{Wu} possibly including the slip length. 

A simple tool to study the effect of polydispersity on polymer slip along low energy substrates is the dewetting experiment.\cite{PGGtext, Gabriele, Vilmin} In this experiment, a hole is formed in a film ca. 100 nm thick, and grows due to a resulting reduction in the capillary energy. Recent theoretical and experimental studies have focused on the detailed morphology of the rim collected ahead of the hole.\cite{Fetzer07L, Fetzer05}  These investigations showed that the boundary condition at the polymer/wall interface directly governs the rim morphology of the dewetting front.\cite{Fetzer07L, Fetzer05} Specifically, the relaxation of the front into the undisturbed portion of the film is characterized with one or two relaxation length scales.  Using these length scales, the slip length is derived through the use of hydrodynamic models incorporating a linear Navier slip boundary condition. The technique, here referred to as rim profile analysis, is a reliable method of slip length determination using dewetting experiments. 

In the present work, we investigate the effect of short chains on slip of highly entangled polystyrene (PS) mixtures on fluorinated substrates as a preliminary step towards understanding the more complex subject of polydispersity. To this end, thin PS films dewetting from fluoropolymer (AF 2400) coated Si wafers are examined. Binary and ternary mixtures of long and short chains with different compositions are studied in a systematic manner. Results show that short chains of molecular weights, $M$, below the polymer critical molecular weight for entanglements, \mc, play an important role in slip of polymer mixtures near low energy surfaces. Among mixtures of the same weight average molecular weight, \mw,  (given that the longest chains are the same among all mixtures), mixtures containing chains of $M<\mc$ exhibit larger slip lengths than those containing short chains of $M>\mc$. This is attributed to physical or chemical disparities between end groups and backbone units, whose presence may also induce dynamical heterogeneity reflected in the slip length. The implication is that PS melts enriched with end groups exhibit a lower effective friction coefficient on fluorinated surfaces than melts with a higher concentration of backbone units. 

\section{Experimental Methods} 
We used atactic polystyrene (Polymer Source Inc.) with $\mw = 5.4, 28.4, 53.3, 86.4, 135.8\textrm{ and }490$\,kg/mol and polydispersity indices below 1.05. In the following, these polymers are referred to as monodisperse polystyrenes with short-hand notations, e.g., PS(5k). Binary and ternary mixtures of different weight fractions were prepared from monodisperse samples. 

In all our mixtures, PS(490k) was used as the primary high molecular weight component (HMW); other polymers were treated as the low molecular weight components (LMW). Several sample sets were examined, the precise details of which are shown in Table 1: 
\begin{itemize}
\item \sampt{S1}{m}: monodisperse PS samples.  
\item \sampt{S2}{fw}: binary and ternary mixtures of a fixed $\mw= 318 \pm 2$\,kg/mol and varying number average molecular weight, \mn.
\item \sampt{S3}{fc}: binary mixtures of a fixed content of HMW ($\phi_\textrm{PS}(490\textrm{k}) = 90$ wt\%) and varying LMW -- this sample set is denoted by e.g. \sampt{S3}{fc}(5k)  and has an approximately constant weight average molecular weight,  expressed as $\mw = 447 \pm 5$\,kg/mol. We note that this spread of  \mw\ is similar to the variation in $M$ expected for pure components (e.g., the PS(490k) used here). 
\item \sampt{S4}{bi}(28k): binary mixtures of PS(28k) and PS(490k) with different compositions. 
\item \sampt{S5}{bi}(86k): binary mixtures of PS(86k) and PS(490k) with different compositions. 
\end{itemize}

\begin{table*}
\caption{Polymer mixtures and dewetting temperatures examined in this work\footnote{Units of \mn\ and \mw\ are in kg/mol.}}
\begin{tabular}{c c || c c c c c c c c c}
\hline 

Sample set & Sym. & $T_\textrm{d}\,(^\circ\textrm{C})$ & $\phi_\textrm{5k}$ & $\phi_\textrm{28k}$ & $\phi_\textrm{53k}$ & $\phi_\textrm{86k}$ & $\phi_\textrm{136k}$ & $\phi_\textrm{490k}$ & \mn\ & \mw \\

\hline 

& PS(5k) &	110 &	1 &	- &	- &	- &	- &	- &	4.7 & 5.4 \\
& PS(28k) &	120 &	- &	1 &	- &	- &	- &	- &	27 &	28.4 \\
\sampt{S1}{m} & PS(53k) &	140 &	- &	- &	1 &	- &	- &	- &	51 &	53.3 \\
monodisperse PS & PS(86k) &	140 &	- &	- &	- &	1 &	- &	- &	85.5 &	86.4 \\
& PS(136k) &	140 &	- &	- &	- &	- &	1 &	- &	 130 &	135.8 \\
& PS(490k) &	150 &	- &	- &	- &	- &	- &	1 &	465 &	490 \\
\hline 
\sampt{S2}{fw}:  &	\sampt{S2}{fw}(20k) &	130/140/150 &	0.22 &	- &	- &	- &	0.19 &	0.59 &	20 &	316 \\
\mw\ fixed and varying \mn &	\sampt{S2}{fw}(86k) &	130/140/150 &	- &	0.25 &	- &	- &	0.14 &	0.60 &	86 &	320 \\
&	\sampt{S2}{fw}(206k) &	130/140/150 &	- &	- &	- &	- &	0.49 &	0.51 &	206 &	316 \\
\hline 
 &	\sampt{S3}{fc}(5k) &	150 &	0.10 &	- &	- &	- &	- &	0.90 &	43 &	442 \\
 \sampt{S3}{fc}: &	\sampt{S3}{fc}(28k) &	150 &	- &	0.10 &	- &	- &	- &	0.90 &	177 &	444 \\
PS(LMW)-PS(490k) &	\sampt{S3}{fc}(53k) &	150 &	- &	- &	0.10 &	- &	- &	0.90 &	257 &	446 \\
 $\phi_\textrm{PS(490k)}= 0.90$ &	\sampt{S3}{fc}(86k) &	150 &	- &	- &	- &	0.10 &	- &	0.90 &	322 &	450 \\
 &	\sampt{S3}{fc}(136k) &	150 &	- &	- &	- &	- &	0.10 &	0.90 &	370 &	455 \\
 \hline 
 &	\sampt{S4}{bi}(28k)-70\% &	130 &	- &	0.70 &	- &	- &	- &	0.30 &	38 &	167 \\
\sampt{S4}{bi}:  &	\sampt{S4}{bi}(28k)-50\% &	130 &	- &	0.50 &	- &	- &	- &	0.50 &	51 &	259 \\
PS(28k)-PS(490k) &	\sampt{S4}{bi}(28k)-30\% &	140 &	- &	0.30 &	- &	- &	- &	0.70 &	79 &	352 \\
 &	\sampt{S4}{bi}(28k)-10\% &	150 &	- &	0.10 &	- &	- &	- &	0.90 &	177 &	444 \\
\hline 
\sampt{S5}{bi}:  &	\sampt{S5}{bi}(86k)-70\%	 &140 &	- &	- &	- &	0.70 &	- &	0.30 &	113 &	208 \\
PS(86k)-PS(490k) &	\sampt{S5}{bi}(86k)-50\% &	140 &	- &	- &	- &	0.50 &	- &	0.50 &	144 &	288 \\
 &	\sampt{S5}{bi}(86k)-30\%	 & 140 &	- &	- &	- &	0.30 &	- &	0.70 &	199 &	369 \\
 &	\sampt{S5}{bi}(86k)-10\% &	150 &	- &	- &	- &	0.10 &	- &	0.90 &	322 &	450 \\
 \hline
\end{tabular}
\end{table*}

To obtain homogeneous solutions for spin coating, samples were dissolved in toluene: the solution was stirred for one day at room temperature and remained on shelf for two additional days to ensure homogeneity. The polymer concentration in toluene for each sample was different depending on the molecular weight (max. $\simeq 25$\,mg/ml) in order to obtain films of undisturbed thicknesses, $h_0$, in the range of  100 -- 200 nm when solutions were spin coated onto freshly cleaved mica sheets.
\begin{figure}[t!]
 \centering
\includegraphics[width=0.75\columnwidth]{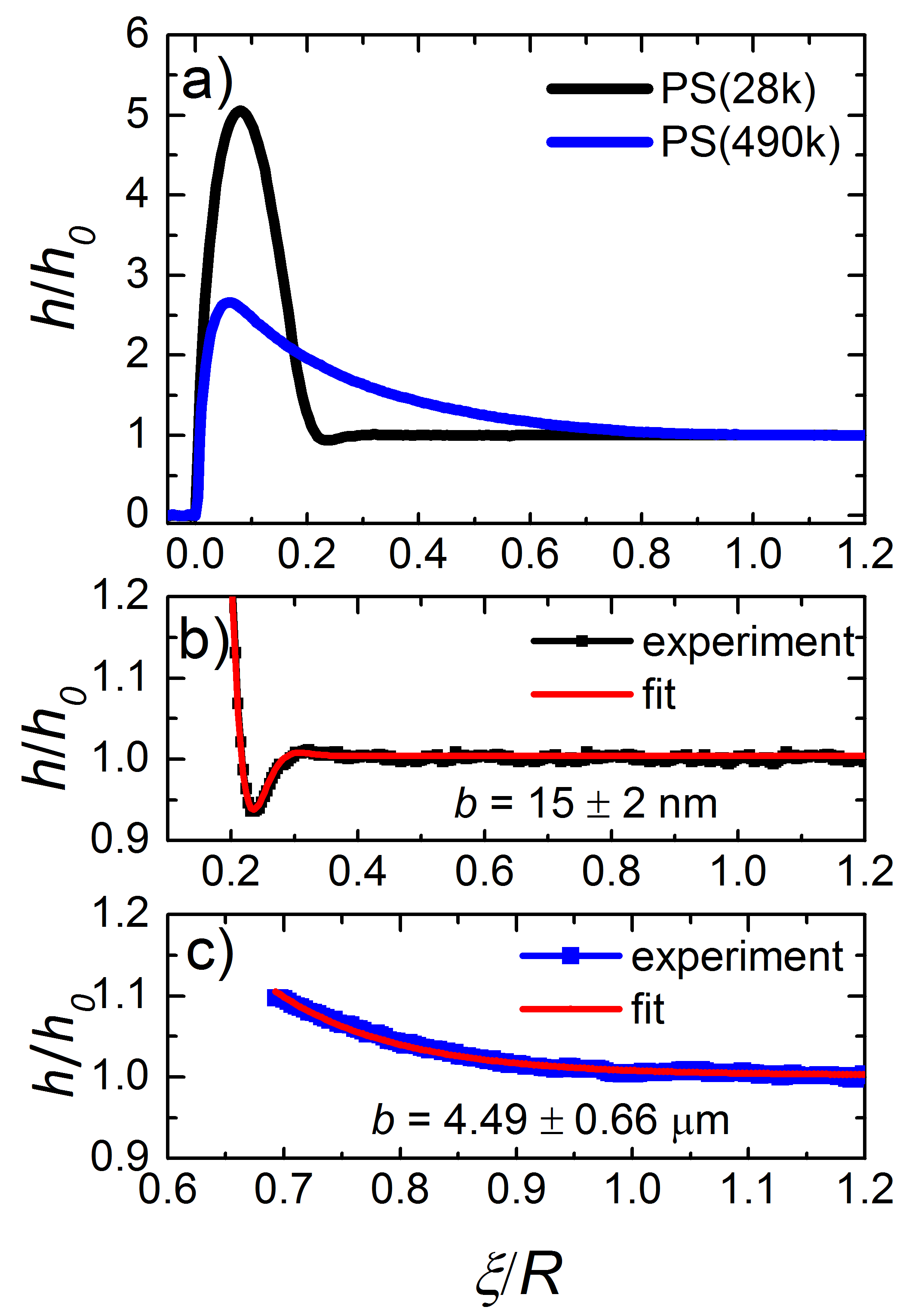}
 \caption{a) Normalized height profiles, $h/h_0$, as a function of distance from the contact line, $\zeta/R$ with $\zeta$ the distance from the contact line and $R$ the hole radius. The profiles shown are for PS dewetting from AF 2400: PS(28k) with $h_0 = 120$\,nm, hole radius $R = 5.9\,\upmu$m and $T = 120\,^\circ$C, and PS(490k) with $h_0 = 172$\,nm, hole radius $R = 6.8\,\upmu$m and $T = 150\,^\circ$C.  b) Best fit to experimental data for PS(28k); and c) the best fit to experimental data for PS(490k) using oscillatory and monotonic decays, respectively.  \label{REXPT}}
\end{figure}

All films spin coated on mica were pre-annealed in air (Linkam) at 150\,$^\circ$C for 1 hr which serves to remove residual stress and solvent. The films were then floated onto the surface of  an ultra-clean water bath (TKA-GenPure, 18.2 M$\Omega$ cm, total organic carbon content $< 5$\,ppb) after which they were picked up from the air/PS surface\cite{SabzevariJP} using hydrophobized Si wafers (Si-Mat Silicon Materials, 100 crystal orientation). Hydrophobization was accomplished using the fluoropolymer AF 2400 (Poly[4,5-difluoro-2,2-bis(trifluoromethyl)-1,3-dioxide-co-tetrafluoro-ethylene], Sigma-Aldrich) with a coating thickness of $h_0 = 15$\,nm. Such a surface treatment leads to a surface energy of ca. 15 mJ/m$^2$.\cite{Baumchen12} 

Samples were heated to above their glass transition temperature, $T_\textrm{g}$,  (for PS: $T_\textrm{g} \leq 100\,^\circ$C) to start dewetting by the nucleation and growth mechanism. Optical microscopy (Leitz) and AFM (Multimode) were used to capture flow dynamics and rim morphology of the dewetting holes. In all experiments, only holes growing with at least a 10 min delay after the dewetting temperature ($110 \leq T_\textrm{d} \leq 150\,^\circ$C) was reached were examined. This can be considered as additional annealing of films on the hydrophobized substrate allowing films to be less affected by frozen-in stresses.\cite{reiter}

\section{Results and Discussion} 
Characterization of slip length, $b$, using the rim profile analysis of the wet side of dewetting holes has previously been well established.\cite{Fetzer07L} Here, we use the third-order Taylor expansion of the full Stokes model developed by Fetzer et al.\cite{Fetzer07L, Baumchen12} to extract the slip length. Examples of the fitting procedure using the rim profile analysis for two monodisperse samples PS(28k) and PS(490k)  are shown in Figure 1, with normalized data obtained using AFM shown in Figure 1a). Figures 1b) and 1c) show the portion of the height profile used for curve fitting as well as the best fit to the experimental data of PS(28k) and PS(490k), respectively. As seen in this figure, PS(490k) exhibits a monotonic decay with a long decay length indicating a strong slip boundary condition at the polymer-solid interface with $b = 4.49 \pm 0.66\,\upmu\textrm{m} > 10 h_0$, whereas PS(28k) exhibits a shorter decay length and an oscillatory decay profile indicating a weak slip boundary condition with $b = 15 \pm 2\,\textrm{nm} < 0.1h_0$. 

We first examine the slip length of monodisperse PSs as shown in Figure 2. The molecular weight dependence of the slip length of PS on AF 1600 has previously been investigated;\cite{Baumchen12} we find that PS on AF 2400 exhibits the same slip behavior as it did on AF 1600. For monodisperse PSs, samples with molecular weights below the critical molecular weight: $M < \mc$ ($\mc \approx 35$\,kg/mol for PS)\cite{rubin} exhibit weak slip with almost a constant value of slip length on the order of a few nanometers. The slip length of samples of $M > \mc$ shows a power law dependence\cite{Brochard94LMR} $b \sim M^3$. The small deviation of the data from the theoretical line at the highest molecular weight may be the result of the extreme sensitivity of the friction coefficient of thin films to atomistic differences in the substrate composition, as shown in ref [\cite{Fetzer07L}].
\begin{figure}[t!]
 \centering
\includegraphics[width=0.75\columnwidth]{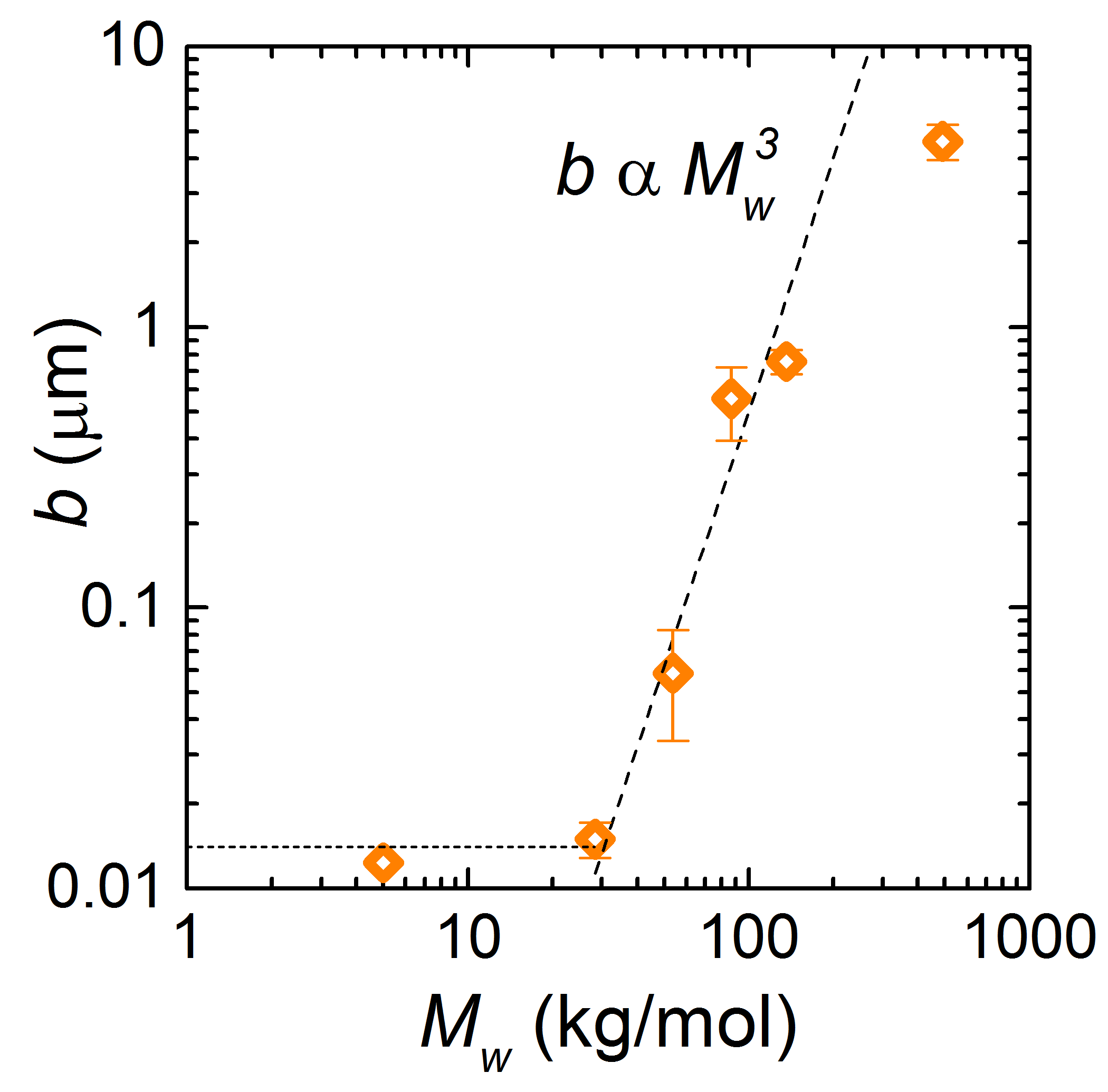}
 \caption{Slip length versus molecular weight of monodisperse PSs on AF 2400. The dashed line shows the theoretical  $M^3$  scaling beginning for $\mw > 35$\,kg/mol. The dotted line drawn for $M < \mc \approx 35$\,kg/mol shows the constant slip length for low molecular weights for which entanglements are not important.  \label{REXPT}}
\end{figure}

In the linear Navier slip model, the slip length is defined as the ratio of melt viscosity, $\eta$, to the friction coefficient, $\kappa$,  i.e., $b = \eta/\kappa$, which is obtained by balancing the stress at the fluid-solid interface.\cite{Brochard94LMR} In the dewetting experiments, the Weissenberg numbers are always very small ($\textrm{Wi} = \gamma\tau < 0.2$, where $\gamma$ and $\tau$ are the shear rate and polymer relaxation time)\cite{Baumchen12} and therefore non-Newtonian behavior such as shear thinning or viscoelasticity\cite{Vilmin} can be safely excluded. The melt viscosity can therefore be represented by the zero shear viscosity, $\eta_0$, which only depends on the value of \mw.\cite{morrison01TXT} What can be critical in slip of polydisperse polymers is the value of \mn Ð sensitive to the content of short chains -- which can affect the polymer slip behavior via the friction coefficient. We therefore start examining mixtures of a fixed \mw\ and varying \mn. The samples of type \sampt{S2}{fc} (see Table 1) contains three samples with $\mw= 318 \pm 2$\,kg/mol and \mn\ values ranging from 20 to 206 kg/mol.

The temporal evolution of the dewetting holes of the \mw-fixed mixtures on AF 2400 at $T_\textrm{d} = 150\,^\circ$C are shown in Figure 3. As seen therein, the growth rate of the dewetting holes significantly increases when \mn\ is decreased. This indicates that flow dynamics of polydisperse polymers can be extremely sensitive to the value of \mn. The enhanced growth rate of dewetting holes with lower \mn\ is likely due to the combined influence of variations in the viscosity and slip length. Regarding the viscosity, we note that although \mw\ is approximately the same among these samples, $T_\textrm{g}$ can however be different among these samples depending on their compositions. For PS, $T_\textrm{g}$ depends sensitively on \mn\ especially for molecular weights below \mc.\cite{Kajiyama} Therefore, samples containing significant numbers of chains with $M<\mc$ possess a lower $T_\textrm{g}$ and thus a lower viscosity.\cite{morrison01TXT} The faster dynamics of the low \mn\ samples can hence be due to their lower viscosity and it is difficult to conclude about their slip behavior based solely on the hole dynamics alone.\cite{Brochard94LMR, Jacobs, McGraw}  This fact highlights the importance of investigating the slip behavior of these samples through their rim profiles.
\begin{figure}[t!]
 \centering
\includegraphics[width=\columnwidth]{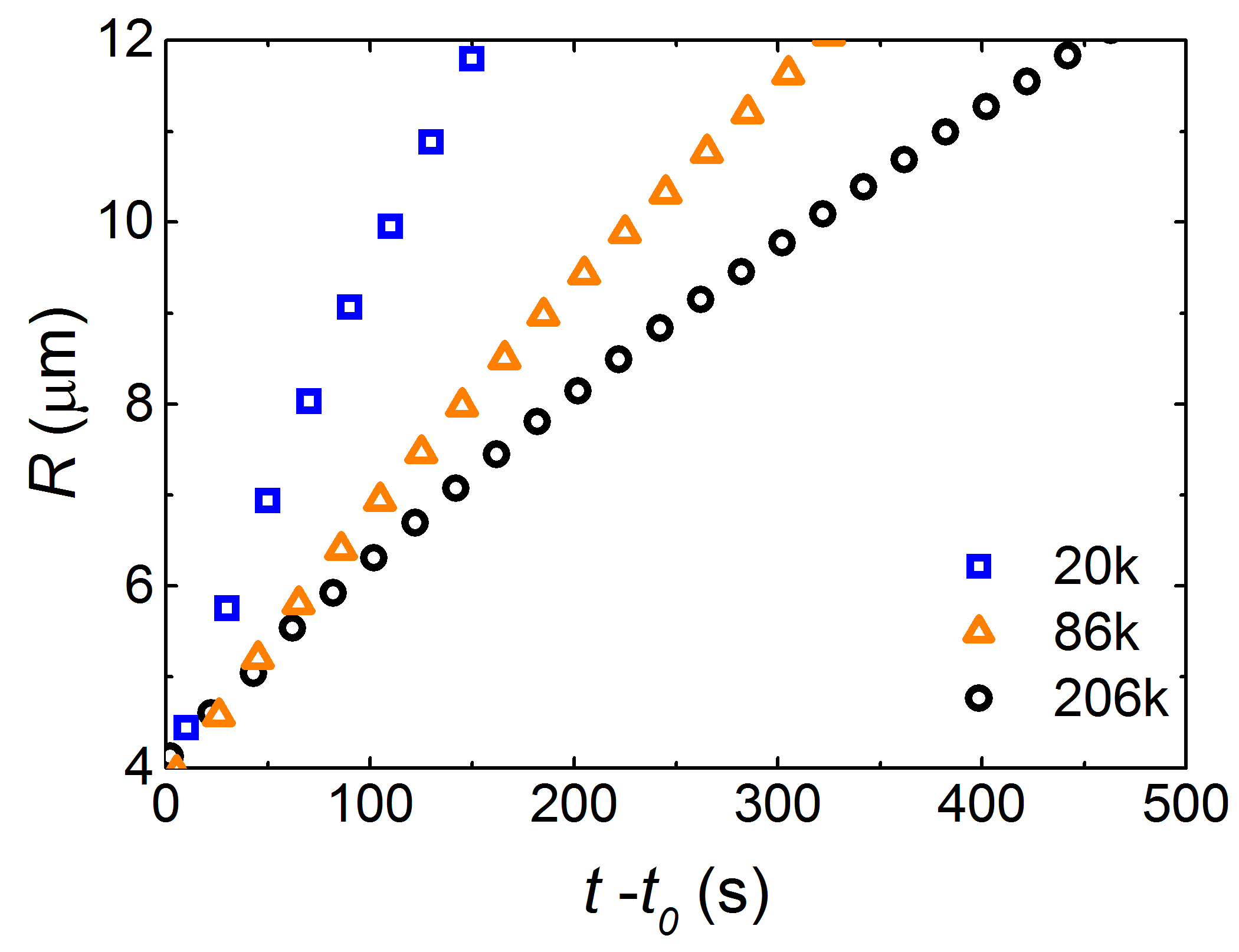}
 \caption{Hole radii, $R$, as a function of time, $t$, for dewetting holes in thin films ($h_0 \approx 150$\,nm) of \mw-fixed mixtures (\sampt{S2}{fw}) on AF 2400 at $T = 150\,^\circ$C; for ease of comparison $t_0$ is taken as the time for which holes have $R = 4 \,\upmu$m. The legend indicates \mn\ of the films.  \label{REXPT}}
\end{figure}

Figure 4 shows the slip length of the same PS samples in Figure 3 along with data at two other dewetting temperatures: 140\,$^\circ$C and 130\,$^\circ$C. Here, all samples of the same \mn\ examined at different dewetting temperatures were prepared from the same film cast on mica, and thus have the same preparation and pre-annealing history. The sample with $\mn = 20$\,kg/mol shows the largest slip length of the three investigated blends of \sampt{S2}{fw}, significantly higher than that of the sample with $\mn = 206$\,kg/mol. 

A decrease of the dewetting temperature has a minimal effect on the value of the slip length of the investigated samples. According to the WLF equation,\cite{Williams} a decrease in the temperature results in an exponential increase in the film viscosity (roughly a decade per ten degrees centigrade for the temperatures presented here). With the exponential increase in the melt viscosity, the slip length might be expected to increase since the linear Navier model used here predicts that $b= \eta/\kappa$. However, consistent with B\"aumchen et al.\cite{Baumchen12} for the slip of PS on AF 1600 (a similar fluoropolymer coating), we observe that the slip length data in Figure 4 are almost insensitive to the change of the dewetting temperature. This is perhaps due to the simultaneous change of the friction coefficient at the PS/PTFE interface along with the PS bulk viscosity. 

The enhanced slip of the low \mn\ samples in Figure 4 signifies the crucial effect of short chains on slip of polydisperse polymers on low surface energy substrates. Since the highest slip length in this graph corresponds to the sample containing the shortest chains as listed in Table 1, it is interesting to further investigate the effect of the size of short chains. To this end, we examine binary mixtures of long and short chains in which the size and content of the HMW component is fixed while the size of LMW component varies (\sampt{S3}{fc} in Table 1). Figure 5a) shows the slip length of \sampt{S3}{fc} samples as a function of the molecular weight of the LMW component denoted by $M$. All dewetting experiments were done on AF 2400 at $T_\textrm{d} = 150\,^\circ$C. The slip length of pure PS(490k) is added to this graph for comparison. We note that PS(490k) has a significantly higher \mw\ in contrast to an average of $\mw = 447 \pm 5$\,kg/mol for binary mixtures. This causes a significant difference in the slip length of pure PS(490k) versus binary samples: mixtures exhibit a lower slip length due to their lower \mw. However, as seen in Figure 5a), the slip length decrease due to the lower \mw\ is only seen for certain binary mixtures. In contrast, binary mixtures containing short chains of $M<\mc$ (samples: \sampt{S3}{fc}(5k) and \sampt{S3}{fc}(28k)) exhibit enhanced slip. 
\begin{figure}[b!]
 \centering
\includegraphics[width=0.75\columnwidth]{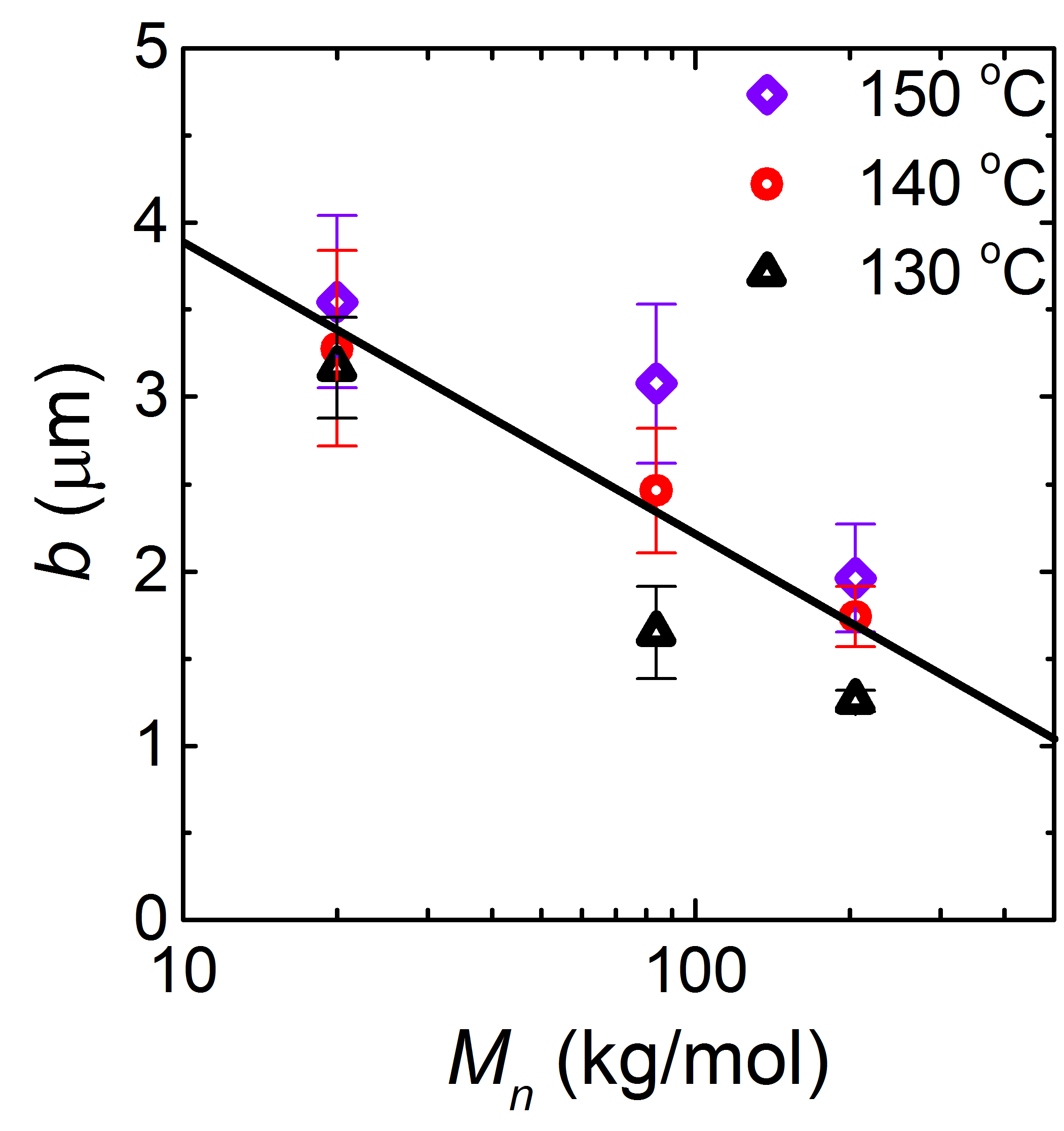}
 \caption{Slip length of \mw-fixed mixtures (\sampt{S2}{fw}) on AF 2400 as a function of \mn\ at different dewetting temperatures, $T_\textrm{d}$, as indicated in the legend. The solid line is a guide to the eye.  \label{REXPT}}
\end{figure}

The slip length of the two previous sample sets, i.e., mixtures of $\mw = 318 \pm 2$\,kg/mol ($T_\textrm{d} = 150\,^\circ$C) and \sampt{S3}{fc} binary mixtures with $\mw= 447 \pm 5$\,kg/mol can be compared. Figure 5b) shows the slip length of the two samples sets as a function of \mn. As can be seen, there is a general trend of the increase of the slip length with the decrease of \mn. These results suggest that among mixtures of the same \mw\ those with a higher content of short chains of $M<\mc$ are likely to exhibit larger slip lengths. 
\begin{figure}[b!]
 \centering
\begin{minipage}{\columnwidth}%
	 \includegraphics[width=0.75\columnwidth]{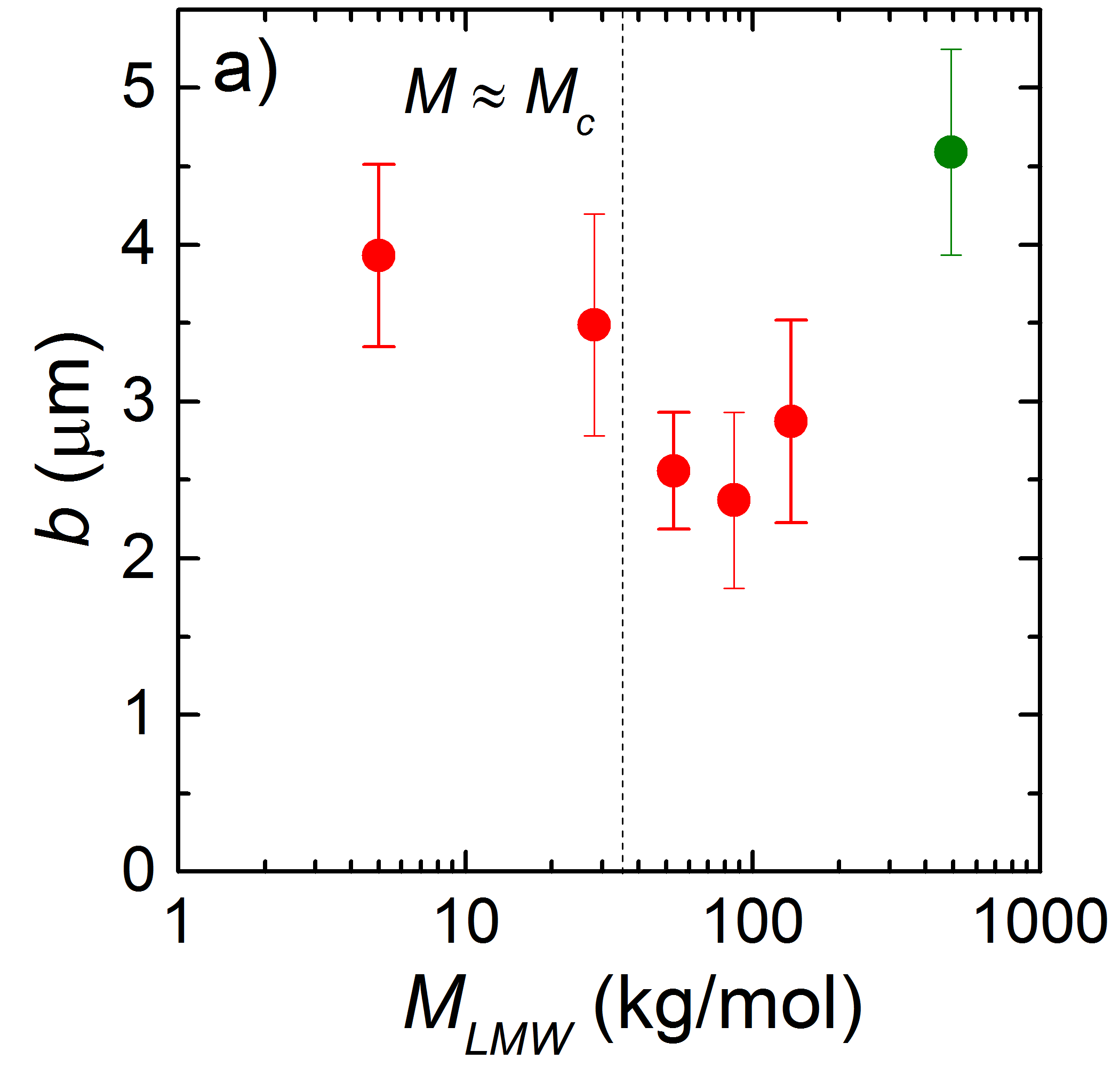}
\end{minipage}%

\begin{minipage}{\columnwidth}
	 \includegraphics[width=0.75\columnwidth]{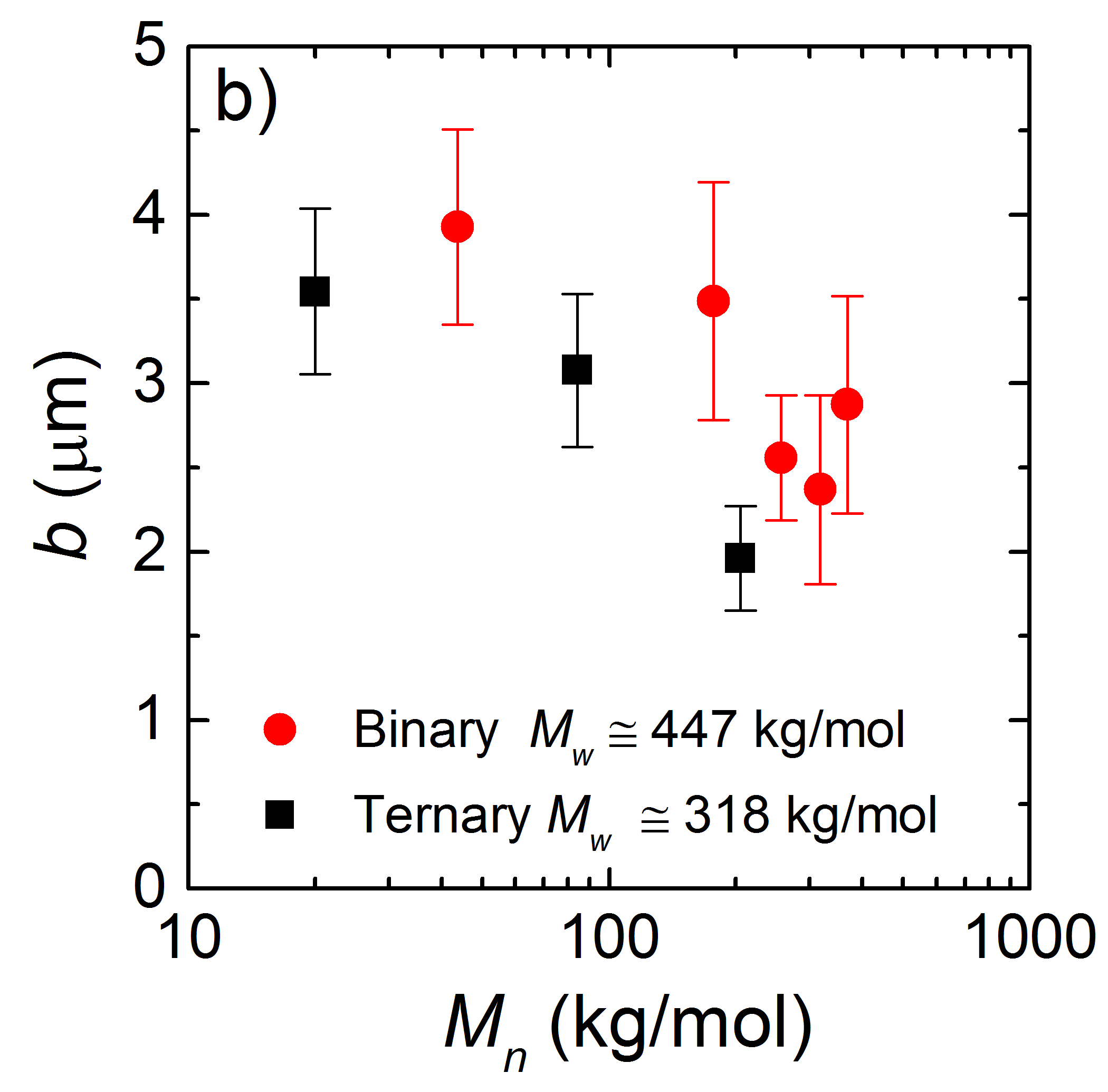}
\end{minipage}%
 \caption{a) Slip length of \sampt{S3}{fc} binary mixtures ($\phi_\textrm{PS}\textrm{(490k)} = 0.90$ in all samples) on AF 2400 versus the molecular weight of the LMW component, $M$; the dotted line shows $M\approx\mc$. b) Slip length versus \mn\ for two sets of \mw-fixed mixtures.  \label{REXPT}}
\end{figure}

To study this hypothesis rigorously, we investigate the effect of the content of short chains in binary mixtures. Two sets of binary mixtures -- \sampt{S4}{bi}(28k) and \sampt{S5}{bi}(86k) detailed in Table 1 -- with different compositions were examined. The choice of PS(28k): $M<\mc$ versus PS(86k): $M>\mc$ is made based on their slip behavior in Figure 5a) and the fact that their molecular weights fall just below and just above \mc, respectively. Figure 6 shows the slip length of \sampt{S4}{bi}(28k)  and \sampt{S5}{bi}(86k) as a function of the composition of PS(490k), $\phi_\textrm{PS}(490\textrm{k})$. In this graph, $\phi_\textrm{PS}(490\textrm{k}) = 1$ represents the slip length of pure PS(490k) and $\phi_\textrm{PS}(490\textrm{k}) = 0$ represents the slip length of pure PS(28k) and PS(86k) samples.  

Interestingly, the slip behavior of the two sample sets relative to one another may be divided into two different regions with respect to $\phi_\textrm{PS}(490\textrm{k}) = 0.5$. For samples in which $\phi_\textrm{PS}(490\textrm{k}) \leq 0.5$, the slip length of mixtures changes according to their corresponding \mw, i.e., binary mixtures containing PS(86k) exhibit equal or larger slip length than those containing PS(28k) because of higher \mw. For samples with $\phi_\textrm{PS}(490\textrm{k}) > 0.5$, the slip length is significantly dependent on the size and content of short chains. Binary mixtures containing 10 and 30 wt\% of PS(28k) exhibit larger slip lengths than those containing PS(86k). In particular, it is interesting to note that the sample \sampt{S4}{bi}(28k)-30\% with $\mw = 352$\,kg/mol and $\mn = 79$\,kg/mol exhibits as large slip length as pure PS(490k). 

The slip behavior of binary mixtures in Figure 6 indicates that both size and content of short chains are important factors in slip of polymer mixtures on low energy surfaces. In this regard, chains of $M < \mc$ are of specific importance when mixed with highly entangled polymers in lower proportions. 

As shown by Brochard and de Gennes in their model for slip of monodisperse entangled chains,\cite{Brochard94LMR} the polymer slip length -- defined as the ratio of melt viscosity to the friction coefficient at the slip plane, $b = \eta/\kappa$ -- shows a strong molecular weight dependence as $b \sim M^3$. This power law dependence is confirmed here in Figure 1, and was first demonstrated experimentally in the work of B\"aumchen and co-workers.\cite{Baumchen12} The molecular weight dependence of $b$ in Brochard and de Gennes relation comes from the dependence of the viscosity in the reptation model, the viscosity scaling with molecular weight\cite{Brochard94LMR} as $\eta \sim M^3$. The friction coefficient in the Borchard-de Gennes relation is $\kappa = \kappa_0$, where $\kappa_0$ is considered to be the constant friction coefficient of main chain monomers, and therefore independent of $M$. While this is a good approximation for monodisperse, highly entangled chains for which this model was developed, it may not be a good approximation for the polydisperse melts we study here. 
\begin{figure}[b!]
 \centering
\includegraphics[width=0.75\columnwidth]{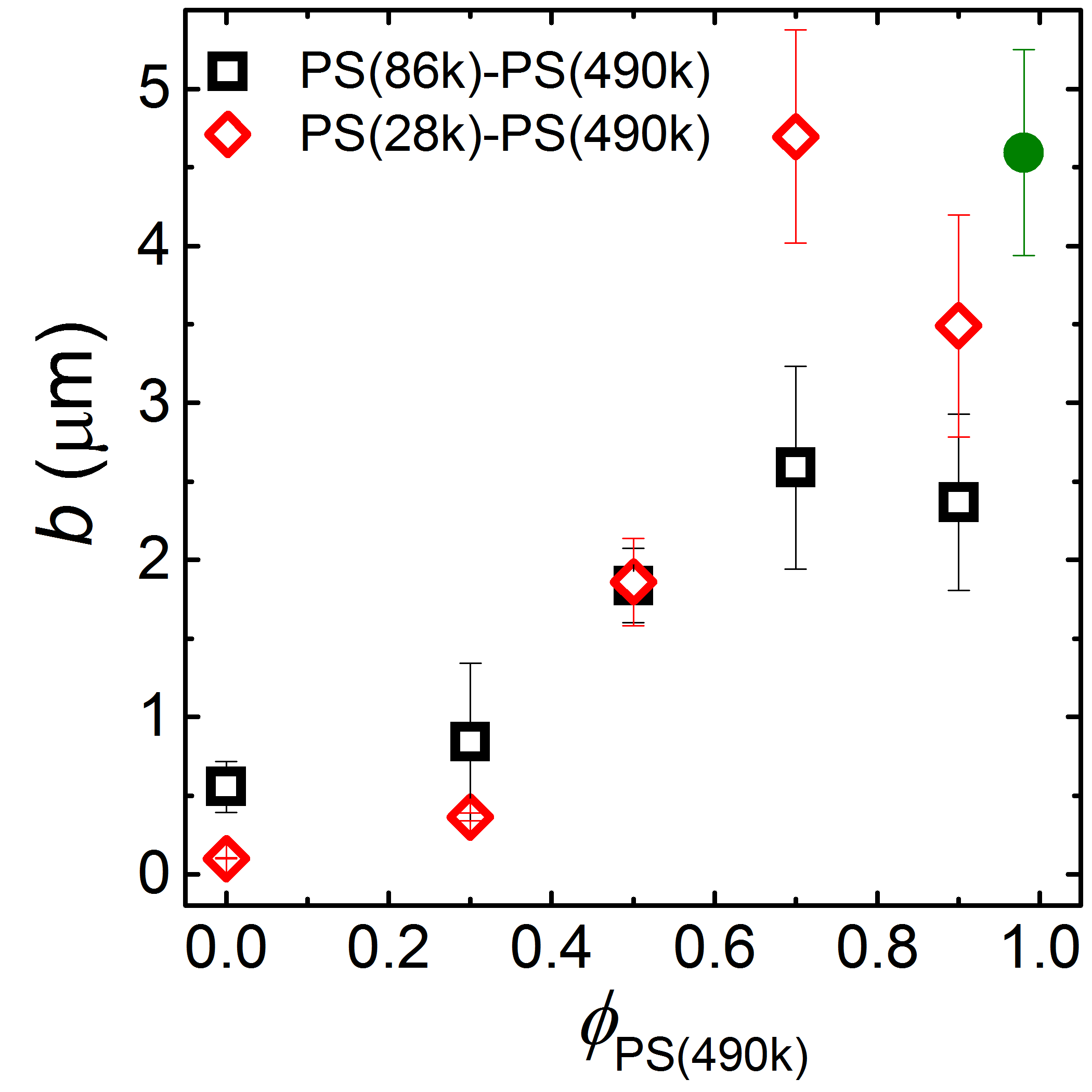}
 \caption{Slip length $b$ versus $\phi_\textrm{PS}\textrm{(490k)}$ for two sets of binary mixtures on AF 2400.  \label{REXPT}}
\end{figure}

The slip mechanism of polymers on low energy surfaces is believed to be due to adhesive failure in which the slip plane is located at the polymer/solid interface. The friction coefficient at the polymer/solid interface can thus be molecular weight dependent due to the change in the content of species (end groups versus backbone units) coming into contact with the solid surface as $M$ varies. As with the surface tension of polymers,\cite{Dee, Jalbert} the friction coefficient is  an interfacial property and may hence depend on the chemistry of monomers present at the solid/liquid interface. Moreover, it has been suggested that the friction coefficient is highly dependent\cite{SabzevariJP} on physical characteristics of chains present at the interface. At small values of $M$, therefore, we expect that the friction coefficient to be significantly affected by the end group properties whereas backbone properties become dominant as $M$ increases (cf. the model of Brochard and de Gennes\cite{BrochardPGG, Brochard94LMR}). 

The disparities between backbone units and end groups are specifically important in polydisperse polymers of broad molecular weight distributions. For instance, polydisperse samples of the same \mw\ differ in the end group to backbone ratio depending on their \mn\ values. Accordingly, polydisperse polymers with large polydispersity indices are rich in the number of end groups which affect their interfacial properties, as revealed by measurements of the surface tension,\cite{Dee} for example, or measurements of chain end segregation near high energy surfaces.\cite{Hariharan}  As we observe in the present work, short chains of $M < \mc$ can significantly enhance the slip length of highly entangled PS mixtures on AF 2400.  This may be due to the presence of end groups at the PS-solid interface via the incorporation of short chains into the sample. As a result of this incorporation, the friction coefficient which is dominantly governed by the backbone characteristics in case of pure PS(490k) becomes influenced by characteristics of end groups in binary mixtures. In addition to the fact that the monomeric friction coefficients may differ between backbone and chain end monomers, full chain dynamics are as well influenced by chain end enrichment.\cite{Hariharan} Chain dynamics of polymer blends are known to show complex behavior on different timescales\cite{Park, Auhl} in bulk, and the extent to which the end-enriched interfacial sample portions are dynamically different from the bulk may be reflected in the space- and time-averaged measurement of the slip length that we make here. We can therefore infer that, for PS, a higher concentration of end groups at the interface more strongly lowers the effective friction coefficient, comprising static and dynamic effects, than it does the viscosity. Consequently, the slip length as extracted from the rim shape analysis is enhanced. 

The extent of the presence of end groups at the interface can also be significantly higher than that in the bulk. There have been many experimental and theoretical studies on surface enrichment of end groups at the surface or interface.\cite{Wu, Tanaka1, Tanaka2, Minnikanti} Molecular dynamics (MD) simulations from Daoulas and coworkers\cite{Daoulas} showed that polyethylene chain ends segregate to both liquid/vapour and solid/liquid interface.  Matsen and Mahmoudi using self-consistent field theory showed that chain ends segregate to a narrow region next to the surface due to entropic effects.\cite{Matsen} In case of polystyrene, Kajiyama et al. \cite{Tanaka1} reported that thin films of PS with $M < 30$\,kg/mol spin coated on silicon wafer show enhanced mobility at the air/PS interface, even at room temperature. The enhanced mobility was attributed to the surface localization of chain end groups. They also showed that surfaces of binary and ternary mixtures of monodisperse PSs show enhanced mobility when a component with $M < 30$\,kg/mol is present.\cite{Tanaka2}. The slip length enhancing effect we observe in our binary and ternary mixtures containing short chains of $M < \mc$ can therefore be as well influenced by the surface enrichment of chain end groups. 

\section{Conclusions}

Slip lengths of monodisperse PSs and their binary and ternary mixtures were examined on AF 2400 using dewetting experiments. It was shown that chains of $M<\mc$ can significantly enhance the slip length of highly entangled PS mixtures on AF 2400: among mixtures of the same \mw those containing chains of $M<\mc$ exhibit larger slip lengths, which is consistent with a lower value of \mn. The slip enhancement effect of chains of $M<\mc$ in their binary mixtures with highly entangled chains is only applicable when the content of the long chain component, $\phi_\textrm{L}$, is dominant ($\phi_\textrm{L} > 0.5$). In particular, we showed that a binary mixture of PS(490k) and PS(28k) containing 30 wt\% PS(28k) exhibits as large slip length as pure PS(490k); the slip length reduced by almost a factor of two when PS(28k) was replaced with PS(86k). 

The present results suggest that the effect of short chains, with $M<\mc$, on slip  of highly entangled PSs on AF 2400, is dominated by the presence of chain end groups at the interface. It can therefore be concluded that, for highly entangled PS, a higher concentration of end groups at the interface results in a lower effective friction coefficient which consequently enhances the slip length. In this regard, the observed slip behavior can be significantly influenced by the surface segregation mechanisms enriching the chain end groups at the interface.

\begin{acknowledgments}
The authors gratefully acknowledge the financial support of NSERC of Canada. JDM was additionally supported by the DFG and by LabEX ENS-ICFP: 
ANR-10-LABX-0010/ANR-10-IDEX-0001-02 PSL. We also thank Prof. Dr. Karin Jacobs for fruitful discussions.
\end{acknowledgments}


\begin{thebibliography}{10}

\bibitem{archer05TXT}
L. A. Archer, in {\em Polymer Processing Instabilities Control and Understanding}, eds. S. G. Hatzikiriakos and K. B. Migler, Marcel Dekker, New York, 2005, {\bf 4}, 73--119.

\bibitem{morrison01TXT}
F. A. Morrison, {\em Understanding Rheology}, Oxford University Press, New York, 2001.

\bibitem{legrand98jr}
F. Legrand, J. M. Piau and H. Hervet, {\em J. Rheol.}, 1998, {\bf 42}, 1389--1402.

\bibitem{Ghanta}
V. G Ghanta, B. L. Riise and M. M. Denn, {\em J. Rheol.}, 1999, {\bf 43}, 435--442.

\bibitem{PGG85rmp}
P. G. de Gennes, {\em Rev. Mod. Phys.}, 1985, {\bf 57}, 828--862.

\bibitem{Denn}
M. M. Denn, {\em Annu. Rev. Fluid Mech.}, 2001, {\bf 33}, 265--287.

\bibitem{BrochardPGG}
F. Brochard and P. G. de Gennes, {\em Langmuir}, 1992, {\bf 8}, 3033--3031.

\bibitem{Tchesnokov}
M. A. Tchesnokov, J. Molenaar, J. J. Slot and R. Stepanyan, {\em J. Chem. Phys.}, 2005, {\bf 122}, 214711.

\bibitem{Brochard94LMR}
F. Brochard-Wyart, P.G. de Gennes, H. Hervert and C. Redon, {\em Langmuir}, 1994, {\bf 10}, 1566--1572.

\bibitem{Leger97}
L. Leger, H. Hervet, G. Massey and E. Durliat, {\em J. Phys.: Condens. Matter.}, 1997, {\bf 9}, 7719--7740.

\bibitem{Sabzevari14M}
S. M. Sabzevari, I. Cohen and P. M. Wood-Adams, {\em Macromolecules}, 2014, {\bf 47}, 3154--3160.

\bibitem{Sabzevari14M2}
S. M. Sabzevari, I. Cohen and P. M. Wood-Adams, {\em Macromolecules}, 2014, {\bf 47}, 8033--8040. 

\bibitem{Fetzer07L}
R. Fetzer, A. Muench, B. Wagner, M. Rauscher and K. Jacobs, {\em Langmuir}, 2007, {\bf 23}, 10559--10566.

\bibitem{Baumchen12}
O. BŠumchen, R. Fetzer, M. Klos, M. Lessel, L. Marquant, H. HŠhl and K. Jacobs, {\em J. Phys. Condens. Matter}, 2012, {\bf 24}, 325102.

\bibitem{Hatzikiriakos12}
S. G. Hatzikiriakos, {\em Prog. Polym. Sci.}, 2012, {\bf 37}, 624--643. 

\bibitem{Ansari1}
M. Ansari, S. G. Hatzikiriakos and E. Mitsoulis, {\em J. Non-Newtonian Fluid Mech.}, 2012, {\bf167?168}, 18--29.

\bibitem{Ansari2}
M. Ansari, Y. W. Inn, A. M. Sukhadia, P. J. DesLauriers and S. G. Hatzikiriakos, {\em Polym.}, 2012, {\bf 53}, 4195--4201.

\bibitem{inn}
Y. W. Inn, {\em J. Rheol.}, 2013, {\bf 57}, 393--406. 

\bibitem{Sabzevari3}
19.	S. M. Sabzevari, S. Strandman and P. M. Wood-Adams, {\em Novel Trends in Rheology VI}, 2015, {\bf 1662}, 030003. 

\bibitem{Brochard96MAC}
F. Brochard-Wyart, C. Gay and P .-G. de Gennes, {\em Macromolecules}, 1996, {\bf 29}, 377--382.

\bibitem{Wu}
D.T. Wu, G.H. Fredrickson, J.-P. Carton, A. Ajdari and L. Leibler, {\em J. Polym. Sci. Part B: Polym. Phys.}, 1995, {\bf 33}, 2373--2389. 

\bibitem{Tanaka1}
K. Tanaka, A. Takahara and T. Kajiyama, {\em Macromolecules}, 1997, {\bf 30}, 6626--6632.

\bibitem{Tanaka2}
K. Tanaka, T. Kajiyama, A. Takahara and S. Tasaki, {\em Macromolecules}, 2002, {\bf 35}, 4702--4706.

\bibitem{Minnikanti}
V.S. Minnikanti, Z. Qian and L.A. Archer, {\em J. Chem. Phys.}, 2007, {\bf 126}, 144905.

\bibitem{PGGtext}
P.G. de-Gennes, F. Brochard-Wyart and D. Qu\'er\'e, {\em Capillarity and wetting phenomena: drops, bubbles, pearls, waves}, Springer, Heidelberg, 2004.

\bibitem{Gabriele}
S. Gabriele, P. Damman, S. Sclavons, S. Desprez, S. CoppŽe, G. Reiter, M. Hamieh, S. Al Akhrass, T. Vilmin and E. Rapha'l, {\em J of Polym. Sci. Part B, Physics}, 2006, {\bf 44}, 3022.

\bibitem{Vilmin}
T. Vilmin and E. Rapha'l, {\em Euro Phys. Lett.}, 2005, {\bf 72}, 781.

\bibitem{Fetzer05}
R. Fetzer, K. Jacobs, A. Munch, B. Wagner and T. P. Witelski, {\em Phys. Rev. Lett.}, 2005, {\bf 95}, 127801.

\bibitem{SabzevariJP}
S. M. Sabzevari, J. D. McGraw, K. Jacobs and P. M. Wood-Adams, {\em Polym.}, 2015, {\bf 78}, 202 --207.

\bibitem{reiter}
G. Reiter, in {\em Glass Transition, Dynamics and Heterogeneity of Polymer Thin Films}, ed. T. Kanaya, Springer, Berlin, Heidelberg, 2013, {\bf 2}, 29--63.

\bibitem{rubin}
M. Rubinstein and R. Colby, {\em Polymer Physics}, Oxford University Press, New York, 2003.

\bibitem{Kajiyama}
T. Kajiyama, K. Tanaka and A. Takahara, {\em Macromolecules}, 1997, {\bf 30}, 280--285. 

\bibitem{Jacobs}
K. Jacobs, R. Seemann, G. Schatz and S. Herminghaus, {\em Langmuir}, 1998, {\bf 14}, 4961--4963. 

\bibitem{McGraw}
J. D. McGraw, O. BŠumchen, M. Klos, S. Haefner, M. Lessel, S. Backes and K. Jacobs, {\em Adv. Colloid Interface Sci.}, 2014, {\bf 210}, 13--20. 

\bibitem{Williams}
M. L. Williams, R. F. Landel and J. D. Ferry, {\em J. Am. Chem. Soc.}, 1955, {\bf 77}, 3701--3707. 

\bibitem{Dee}
G. T. Dee and B. B. Sauer, {\em Adv. Phys.}, 1998, {\bf 47}, 161--205. 

\bibitem{Jalbert}
C. Jalbert, J. T. Koberstein, I. Yilgor, P. Gallagher and V. Krukonis, {\em Macromolecules}, 1993, {\bf 26}, 3069--3074.

\bibitem{Hariharan}
A. Hariharan, S. K. Kumar, M. H. Rafailovich, J. Sokolov, X. Zheng, D. H. Duong, S. A. Schwarz and T. P. Russell, {\em J. Chem. Phys.}, 1993, {\bf 99}, 656--663.

\bibitem{Park}
S. J. Park and R. G. Larson, {\em Macromolecules}, 2004, {\bf 37}, 597--604. 

\bibitem{Auhl}
D. Auhl, P. Chambon,  T. C. B. McLeish and D. J. Read, {\em Phys. Rev. Lett.}, 2009, {\bf 103}, 136001. 

\bibitem{Daoulas}
K. Ch. Daoulas, V. A. Harmandaris and V. G. Mavrantas, {\em Macromolecules}, 2005, {\bf 28}, 5780--5795. 

\bibitem{Matsen}
M. W. Matsen and P. Mahmoudi, {\em Eur. Phys. J. E}, 2014, {\bf 37}, 78.

%
\end{thebibliography}
\end{document}